\documentclass[12pt]{article}
\long\def\abst#1{\long\gdef\@abst{#1}}
\long\def\inst#1{\long\gdef\@inst{#1}}
\long\def\kword#1{\long\gdef\@kword{#1}}

\newcommand\mib[1]{\boldsymbol{#1}}
\usepackage{amsmath}
\usepackage{graphicx}

\begin{document}
\begin{quote}
\begin{center}
\large{\textbf{Jarzynski Equality \\for an Energy-Controlled System}}
 \end{center}
 \end{quote}
 
\begin{center}
\small{Hitoshi \textsc{Katsuda}\footnote{E-mail: katsuda@stat.phys.titech.ac.jp} Masayuki \textsc{Ohzeki}$^{2}$}
 \end{center}

\begin{center}
\footnotesize{\textit{$^{1}$Department of Physics, Tokyo Institute of Technology, Oh-okayama, Meguro-ku, Tokyo 152-8551, Japan \\
 $^{2}$ Department of System Science, Graduate School of Informatics, Kyoto University, Yoshida-Honmachi, Sakyo-ku, Kyoto 606-8501, Japan}}
\end{center}

The Jarzynski equality (JE) is known as an exact identity for nonequillibrium systems\cite{originalje}. The JE was originally formulated for isolated and isothermal systems, while Adib reported an JE extended to an isoenergetic process\cite{isoje}. On an isoenergetic process, the functional form of the Hamiltonian is altered with the energy fixed at a constant value $E$ under the effect of an aritificial field term which is added to the Hamilton dynamics. This kind of JE is useful to calculate the entropy which replaces the Helmholtz free energy appearing in the original JE.

On the other hand, the limitation of the constant energy seems to be an obstacle to widely apply the JE to more interdisciplinary problems such as optimization problems. Another problem is that it takes too long time to estimate some physical quantities since the entropy at the different energy values cannot be calculated at the same time with the JE. In the present paper, we extend the JE to an energy-controlled system in order to overcome these difficulties. In our study, we make it possible to control the instantaneous value of the energy arbitrarily in a nonequilibrium process. Under our extension, the new JE is more practical and useful to calculate the number of states and the entropy than the isoenergetic one. Furthermore, we expect that our JE can be used also for applications to computation and information science beyond physics. We will show an instance of possible application of our JE to a kind of optimization problems which is equivalent to physical problems to find the ground state\cite{annealing}.

Following the derivation of Adib, let us consider a classical system which evolves under the following dynamics, where an artificial field term 
$\mib{F}=(\mib{F}_x, \mib{F}_p)$ is added to the Hamilton dynamics,
\begin{equation}
\dot{\mib{x}}=\frac{\partial H}{\partial \mib{p}}+ \mib{F}_x(\mib{\Gamma}),\>\>\>
\dot{\mib{p}}=-\frac{\partial H}{\partial \mib{x}}+ \mib{F}_p(\mib{\Gamma}).
\label{eq:Hamilton dynamics}
\end{equation}
The above $\mib{\Gamma}=(\mib{x},\mib{p})$ describes a point on the phase space. In order to control the value of the energy as a function of time $E(t)$ from $t=0$ to $\tau$, we choose the functional form of 
$\mib{F}=(\mib{F}_x, \mib{F}_p)$ as
\begin{equation}
\mib{F}(\mib{\Gamma}) = \frac{\mib{X}}{\mib{X}\cdot \partial _{\mib{\Gamma}}H}\biggl( \frac{\rm d \it E}{\rm d \it t}-\frac{\partial H}{\partial t}   \biggr),
\label{eq:energy reservoir}
\end{equation}
where $\mib{X}$ is a vector on the phase space satisfying 
$\mib{X}\cdot \partial _{\mib{\Gamma}}H\neq 0$.
From eqs. (\ref{eq:Hamilton dynamics}) and (\ref{eq:energy reservoir}), we immediately recognize that the energy is accurately controlled as $E(t)$ since we have
\begin{equation}
\frac{\rm d \it H}{\rm d \it t}=\frac{\partial H}{\partial \mib{\Gamma}}\cdot \dot{\mib{\Gamma}}+\frac{\partial H}{\partial t}=\frac{\rm d \it E}{\rm d \it t}.
\label{eq:control}
\end{equation}

Under the dynamics (\ref{eq:Hamilton dynamics}), the ensemble density $\rho _t(\mib{\Gamma})$ evolves following the Liouville equation 
\begin{equation}
\rho _t(\mib{\Gamma }_t) = \rho _0(\mib{\Gamma}_0) \rm e^{\it -t \overline{\rm \Lambda _{\it t}}(\mib{\Gamma}_t)},
\label{eq:Liouville eq}
\end{equation}
where
\begin{equation}
\overline{\Lambda _t}(\mib{\Gamma}_t) = \frac{1}{t}\int _0 ^t \rm d \it t' \rm \Lambda \it (\mib{\Gamma }_{t'}).
\label{eq:Lambda bar}
\end{equation}
Equation (\ref{eq:Lambda bar}) is the time average of the ``phase space compression factor" 
$\Lambda (\mib{\Gamma}_t) = \partial _{\mib{\Gamma}_t} \cdot \dot{\mib{\Gamma}_t}$
along the trajectory that connects $\mib{\Gamma}_0$ to $\mib{\Gamma}_t$.\cite{liouville} 
This factor determines the time evolution of the ensemble density $\rho _t(\mib{\Gamma}_t)$ from $t=0$ to $\tau$, and plays the central role to estimate the entropy efficiently.
The system is set to be in an equilibrium state at the initial time $t=0$ similarly to the original JE. Therefore 
$\rho _0(\mib{\Gamma}_0) $ is equal to the microcanonical distribution at $E = E(0)$
\begin{equation}
\rho _0(\mib{\Gamma}_0) =\frac{\delta (H(\mib{\Gamma}_0)-E(0))}{\Omega _0},
\label{eq:inital}
\end{equation}
where $\Omega _0$ is the number of states at $t=0$. 

Let us consider the average of $\rm e^{\tau\overline{\Lambda _{\tau}}}$ over all possible realizations from $t=0$ to $\tau$:
\begin{align}
\langle \rm e^{\it \tau\overline{\Lambda _{\tau}}} \rangle  & = \int \rm d \it \mib{\Gamma}_{\rm \tau}\rho _{\tau}(\mib{\Gamma}_{\tau}) \rm e^{\tau \overline{\Lambda _{\tau}}} \\
  & = \int \rm d  \mib{\Gamma}_{\tau} \frac{\delta (\it H(\mib{\Gamma} _{\rm 0}\it)-E(\rm 0))}{\Omega _0}. 
\label{average}
\end{align}
In the second line, eqs. (\ref{eq:Liouville eq}) and (\ref{eq:inital}) have been used.
Since the value of the Hamiltonian is controlled as $H(\mib{\Gamma}_t) = E(t)$, we have
\begin{equation}
H(\mib{\Gamma}_\tau) -E(\tau)=H(\mib{\Gamma}_0) -E(0).
\label{eq:control}
\end{equation}
Therefore $H(\mib{\Gamma} _0)-E(0)$ in the Dirac delta function of eq. (\ref{average}) 
can be replaced by $H(\mib{\Gamma} _{\tau})-E(\tau)$. Then, eq. (\ref{average}) is reduced to
\begin{equation}
\langle \rm e^{ \tau\overline{\Lambda _{\tau}}} \rangle = \rm e^{ \Delta \it S},
\label{eq:myje}
\end{equation}
where $ \Delta S = \ln (\Omega _{\tau}/\Omega _{0})$ is the entropy difference between the equilibrium states at the different energy values $E(0)$ and $E(\tau)$.
If we estimate the entropy difference with such a naive method as directly calculating the entropy at the initial and final times, we have to make the estimation of entropy twice. On the other hand, our formula (\ref{eq:myje}) enables us to obtain the entropy difference by taking just a single average of 
$\rm e^{\tau \overline{\Lambda _{\tau}}}$ .
The above equality is the our central result, and we will show a practical aspect of our formulation next.

In the application to an optimization problem, let us consider a simple optimization problem to find a minimum of the potential energy. 
If the potential energy has a complex structure, simple searching techniques such as the steepest decent method\cite{sg} does not work well to reach the minimum\cite{np}. Therefore we need to employ an approximate algorithm in practice. 
We show that our JE can be used for a quantitative estimation, which indicates how much a tentative solution given by the approximate approach differs from the actual minimum solution. 

Let us consider an arbitrary form of the potential energy $V(\mib{x})$ of continuous variables $\mib{x}$ which has no degeneracy at the ground state, and assume that an approximate solution with energy $\mathcal{E}$ has been obtained through an approximatiove algorithm. With the theory to control the energy, the energy of the system is changed from $E(0) = \mathcal{E}'$ to $E(\tau) =\mathcal{E}$, where $\mathcal{E}'$ is larger than $\mathcal{E}$. 
As the Hamiltonian $H$ is set as 
\begin{equation}
H=\mib{p}^2+\frac{t}{\tau}V(\mib{x}) ,
\label{eq:Hforopti}
\end{equation}
we can easily calculate the entropy at the initial time $S(0)$ since the system equals to non-interacting particles with the mass of each particles $m=1/2$ at $t=0$. Then, the entropy at the final time $S(\tau) = S(0)+\Delta S$ is obtained with $\Delta S$ estimated in the following way, and the results are independent of the value of $\tau$. First, we randomly choose an initial condition $( \mib{x}_0, \mib{p}_0)$ from the set 
$\{ (\mib{x}$, $\mib{p}) | H(\mib{x}, \mib{p}; t=0) = \mathcal{E}' \}$ and note that the initial Hamiltonian depends only on $\mib{p}$. We obtain the path toward a phase point with the lower energy following the energy-controlled dynamics (\ref{eq:Hamilton dynamics}) under the given initial condition. Some initial conditions result in the divergence of the factor appearing on the right-hand side of eq. (\ref{eq:energy reservoir}) when the system is trapped in a local minimum with $\partial_{\mib{\Gamma}_t}H=0$. Such a divergence means that the energy is not to be decreased to $\mathcal{E}$ and the system evolving from such initial conditions are unable to reach any points on the phase space at $t=\tau$. Therefore such samples should be excluded in taking the average of $\rm e^{\tau \overline{\Lambda _{\tau}}}$.
Finally, we take the average of $\rm e^{\tau \overline{ \Lambda _{\tau}}}$ only in the case of the absence of such divergences, and we obtain $\Delta S$ appearing on the right-hand side of our JE (\ref{eq:myje}).

The minimum value of entropy is $-\infty$ in the case of classical systems.
Therefore $S(\tau) \to -\infty$ implies that the solutions for the given potential energy are close to the minimum point.
In other words, estimating the difference of the entropy can be an indicator how much the solution differs from the actual minimum.

For instance, we estimate the entropy gap $\Delta S$ for the potential 
\begin{equation}
V(x)=x^2-x/2-2\sin (3\pi (x+2))-1/2,
\label{eq:potential}
\end{equation}
where $\tau=10$,
 $\mathcal{E}'=3$, and $\mathcal{E}=2.9, 2.99$ as Fig . $1$. Results are $\Delta S=0.561$ ($\mathcal{E}=2.9$), $0.464$ ($\mathcal{E}=2.99$).
\begin{figure}[tb]
\begin{center}
\includegraphics[scale=0.3]{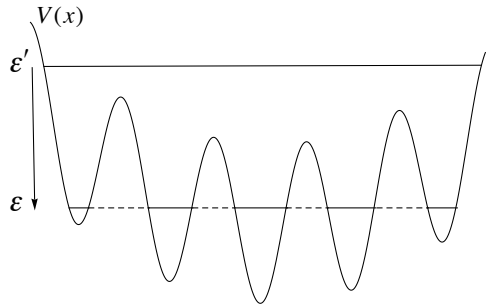}
\end{center}
\caption{The energy is controlled from $\mathcal{E'}$ to $\mathcal{E}$ for the potential energy $V(x)=x^2-x/2-2\sin (3\pi (x+2))-1/2$. Length of the horizontal lines indicating $\mathcal{E}$ and $\mathcal{E}'$ corresponds to the entropy at each energy.}
\label{f1}
\end{figure}
A lower $\mathcal{E}=2.9$ gave a larger entropy difference $0.561$. The difference of $\Delta S$ between 
$0.561$ and $0.464$ is equal to $0.097$, which indicates that the case of $\mathcal{E}=2.9$ is closer to the actual minimum solution than the case of $\mathcal{E}=2.99$ by $0.097$. 

JE (\ref{eq:myje}) can be used also for the calculation of the canonical average of an observable $\mathcal{O}(E)$, which can be expressed as follows,
\begin{equation}
\langle \mathcal{O}(E) \rangle = \frac{\int \rm d \it E \mathcal{O}(E)\rm e^{ \it S(E) -\rm \beta \it E}}{\int \rm d \it E \rm e^{\it S(E) -\rm \beta \it E}},
\label{eq:average}
\end{equation}
where $\rm e^{\it S(E)}$ is equal to the number of states at the energy $E$ and can be obtained from the  JE (\ref{eq:myje}).
While Adib has introduced a similar method with JE for an isoenergetic process\cite{isoje}, such a method takes a long time in practice to estimate
the entropy as a function of $E$. In our method we can control the energy dynamically and therefore obtain the values of $\rm e^{\tau \overline{ \Lambda _{\tau}}}$ at various values of the energy from $E(0)$ to $E(\tau)$ in a single realization.
Then, the necessary time to take the above average can be much shorter than the method of Adib.

In future, we would modify the JE on an energy-controlled process for a quantum system. 
A quantum system is more useful to estimate the efficiency of  an approximate solution of an optimization problem since the entropy at the ground state is equal to $0$ which is clearly more suitable for the quantitative estimation than $- \infty$ of classical systems. Furthermore, the tunneling effect and discrete energy levels seem useful also for estimating the entropy of systems with discrete variables or complex landscape of potential energy like spin systems.

\section*{Acknowledgment}
The authors are grateful to Hidetoshi Nishimori for enlightening discussions and some revisions in a draft.

\end{document}